# Low-cost prediction of molecular and transition state partition functions via machine learning

Evan Komp[1,*] and Stéphanie Valleau[1,*]

[1]*Department of Chemical Engineering, University of Washington, Seattle, Washington 98195, United States*

## Abstract

We have generated an open-source dataset of over 30000 organic chemistry gas phase partition functions. With this data, a machine learning deep neural network estimator was trained to predict partition functions of unknown organic chemistry gas phase transition states. This estimator only relies on reactant and product geometries and partition functions. A second machine learning deep neural network was trained to predict partition functions of chemical species from their geometry. Our models accurately predict the logarithm of test set partition functions with a maximum mean absolute error of 2.7%. Thus, this approach provides a means to reduce the cost of computing reaction rate constants ab initio. The models were also used to compute transition state theory reaction rate constants prefactors and the results were in quantitative agreement with the corresponding ab initio calculations with an accuracy of 98.3% on the log scale.

.

## Introduction

The reaction rate constant or speed of a chemical reaction defines its success. Evaluating reaction rate constants in organic chemistry is critical for drug design, catalysts design, and so forth. Unfortunately, reactions are usually part of large chemical networks. The challenge thus becomes evaluating the kinetics of these networks, i.e. all single reaction rate constants must be computed. The full ab initio calculation of a reaction rate constant can take several months of computational and human time.[1] Indeed, most kinetic theories require the exploration of one or more potential energy surfaces[2–4] and rely on the evaluation of a reactant and transition state partition function.

On the other hand, once input features have been computed, a trained machine learning (ML) estimator may predict chemical properties within seconds.[5–13] Machine learning has successfully been applied to evaluate electronic energies,[14–16] accelerate the search for minimum energy paths,[17–19] and predict kinetic properties[20] such as quantum reaction rate constants for one dimensional minimum energy paths.[21,22]

In this work we investigated the use of ML to predict transition state partition functions without knowledge of its geometry as well as predicting partition functions when geometries are known (Figure 1). We generated a dataset of over 30k partition functions for gas phase organic chemistry species extracted from an existing dataset[23] of reactant, product, and transition state structures for unimolecular reactions. With this data, two deep neural network (DNN) partition function estimators were optimized. The first DNN, "Qest", predicts the natural logarithm of the molecular partition function from featurized molecular geometry and inverse temperature, $1/T$. The second DNN, "QesTS", predicts the natural logarithm of the partition function of an unknown transition state, log $Q_{TS}(T)$, from the difference between product and reactant featurized geometries, the reactant and product partition functions on the log scale, and $1/T$. These models were then used to predict partition functions to be used in the computation of transition state theory (TST) reaction rate constants. Our reaction rate constant predictions were also compared to ab initio TST reaction rate constants. The various steps of our workflow, our results, and conclusions will be discussed in the following subsections.

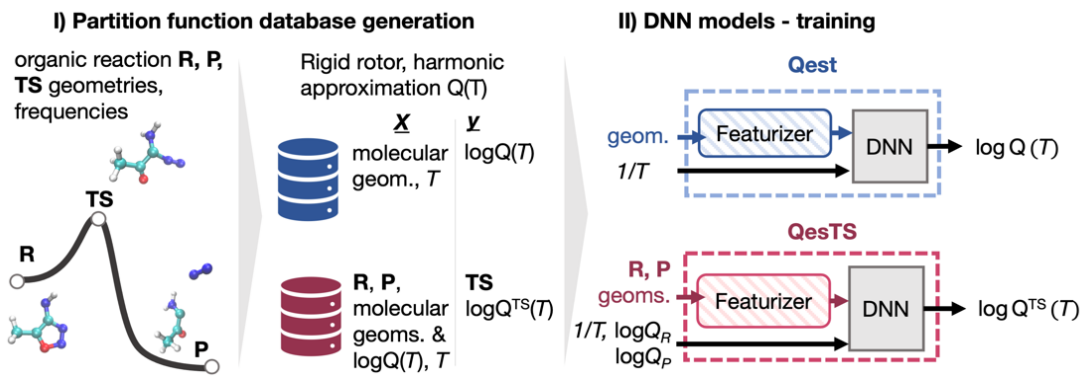

Figure 1: Panel I) Partition functions were computed for a set of DFT optimized reactant, product, and transition state geometries taken from a dataset of 11,961 unimolecular reactions.[23] Panel II) With this data, a deep neural network (DNN), *Qest*, was trained to predict a structure's partition function from its geometry and the inverse temperature, $1/T$. A second DNN, *QesTS*, was trained to predict the partition function of an unknown transition state from the reactant and product structures, partition functions, along with $1/T$.

(*) Corresponding authors: evankomp@uw.edu, valleau@uw.edu

## Results and discussion

### Generation of a partition function dataset

Partition functions were computed for 35,883 reactant, product, and transition state geometries taken from a dataset[23] which contained 11,961 gas phase, organic chemistry reactions involving molecules of at most seven C, O and or N atoms. The partition functions were computed under the assumption that electronic, translational, rotational, and vibrational degrees of freedom were separable (Eq. 1), and by using the rigid-rotor and harmonic-oscillator approximations:

$$Q(T) \approx Q_{el}(T)\, Q_{trans}(T)\, Q_{rot}(T)\, Q_{vib}(T). \quad (1)$$

See Table S1 in the Supporting Information for the equations of each partition function term in Eq. 1. Although some products consisted of two or more distinct molecules, these had been considered as single structures when energies and hessians were computed in the original dataset.[23] We therefore used these single geometries to compute the partition functions. A more accurate representation would require the separation of the product geometry into the individual molecular geometries. When computing the vibrational partition functions, zero-point energies were included. For the rigid rotor rotational partition function, we determined the symmetry number by identifying invariant symmetry operations for each molecule using pymatgen,[24] and by using our own software to determine which symmetry operations were proper. We accounted for the exception of linear molecules for which $C_2$ rotations are returned as improper reflections by pymatgen.

Partition functions were computed over a set of 50 temperatures randomly sampled for each reaction from a uniform distribution of $1/T$ within the range of T = (50, 2000] K. A histogram of the natural logarithm of the partition functions for all structures at all temperatures is shown in Figure 2. The histogram shows a smaller count of low values of the natural logarithm of the partition function. This is due to the dominating contribution of the vibrational partition function to Eq. 1. Our data has been made open source and is available to download from Ref 25.

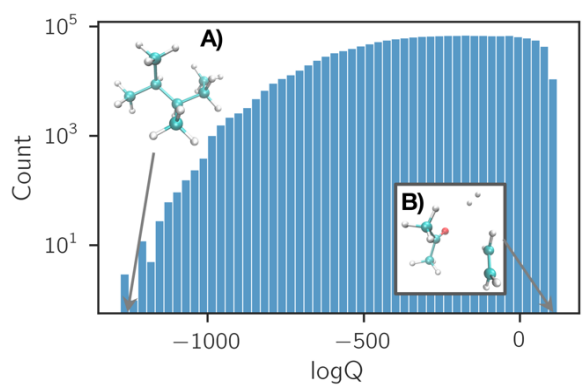

Figure 2: Histogram of all partition function values computed in our dataset for 35,883 organic chemistry species. Each entry is either a reactant, product, or transition state at one of 50 temperatures $T \in (50, 2000]$ K, sampled uniformly with respect to $1/T$. Structures that exhibit the smallest and largest partition functions in the dataset are depicted in subpanels A) and B). The shape of this histogram is largely due to the dominating contribution of the vibrational partition function in Eq. 1.

The entire dataset was split into a hold out test set (10%) and a development set (90%). The hold out set was used to test final ML predictors. The development set was further split into 5 folds to use for cross validation during model optimization. Both the hold-out and the fold splitting were conducted using the scaffold of the reactant molecule.[26] By taking into account molecular structure backbones in scaffold splitting we ensured that the structural overlap of the different datasets was minimized, producing a better estimate of the models accuracy on unseen data. For *Qest*, the input consisted of a molecule's featurized structure (reactant, product, or transition state) and the inverse temperature, while the target was the corresponding natural logarithm of the partition function at that temperature. On the other hand, for *QesTS* the inputs were the difference between product and reactant featurized structures, product and reactant partition functions, and the inverse temperature while the target was the natural logarithm of the transition state partition function.

### Optimization and training of machine learning models

We first determined the optimal format of the input data. This involved a screening of featurization methods and data standardization. We found that the EncodedBonds[27] featurizer with min-max scaling and target normalization were optimal for the *Qest* Model. For the *QesTS* model we employed the same input data feature representation as *Qest*, i.e. Encoded Bonds. We also carried out the screening of data standardization and found it was better not to use standardization. We believe that this is because the distribution of values for the difference in product and reactant feature vectors is peaked around zero; also, the distribution of partition function values for transition states is narrower than the overall partition function distribution for all species. For more information see Supporting Information section 2.

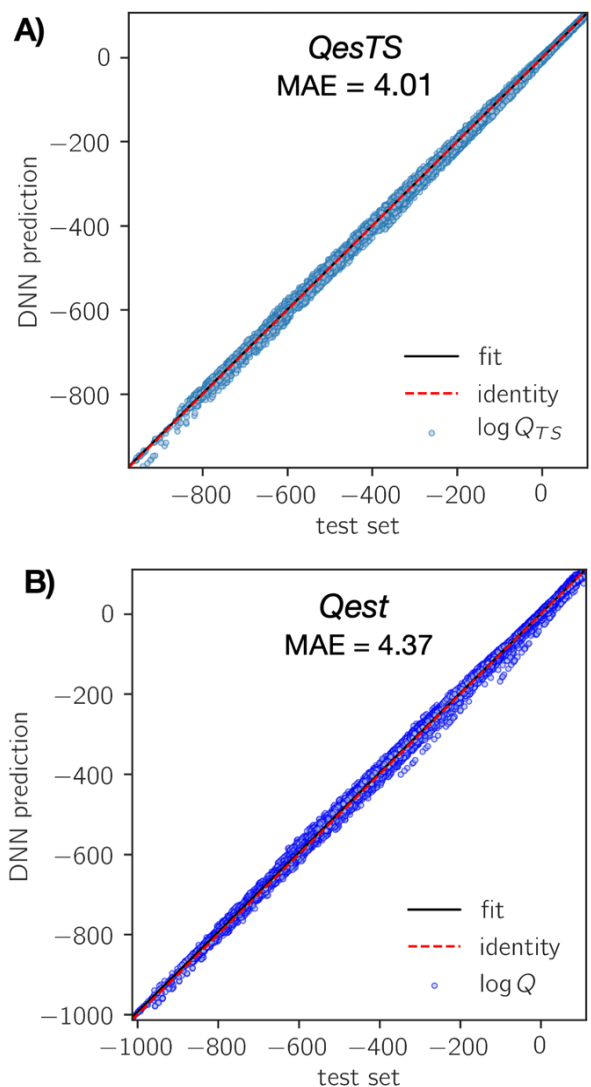

Figure 3: Panel A) Parity plot of the final *QesTS* DNN predicted values of transition state $\log Q_{TS}$ (y-axis) with respect to exact unseen test set values (x-axis). A perfect model is represented by the identity line (red dashes). Predicted examples (blue dots) closely match the true value and the overall MAE on the $\log Q$ is 4.01 (2.0%). Panel B) Same as panel A but for the *Qest* prediction of $\log Q$. Here the overall MAE on the test set $\log Q$ is 4.37 (2.1%). Percent error is compared to the test set standard deviation. We see a strong agreement of our DNN predictions with the test set values.

With these optimal input features, we carried out a search over DNN hyperparameters to identify the best activation function, regularization, bias and weight initialization, learning rate, and neuron configuration for *Qest* and *QesTS*. For more information see Supporting Information section 3.

**Partition function prediction**

The optimal featurizer, standardization, and hyperparameters (Table S4) were used to train the final DNN models on the development dataset and test its performance on the test set. To limit extrapolation in the final model, some outliers were dropped from the total dataset.

In Figure 3 we show parity plots of the final *QestTS* (panel A) and *Qest* (panel B) model predictions with respect to the test set. In both cases we see a low test set mean absolute error (MAE) of 4.01 (2.0%) and 4.37 (2.1%) on the logarithm of the chemical species or transition state partition function for *QesTS* and *Qest* respectively. Percentages are MAE compared to the test set standard deviation. Also, in both cases the spread of the distribution of predictions around the identity is quite narrow. This indicates that both DNN models accurately predict the partition functions.

To verify that the *Qest* model was learning from the molecular structures in the dataset and not simply from the temperature, we created a "null" linear model $\log Q(T) = m \cdot 1/T + b$ and fitted it to the development set. The null model performed poorly on the test set with an MAE of 31.8%. Our *Qest* model error was much lower: 1.9%, confirming that the *Qest* model learned from the molecular structure.

The prediction MAEs on the test set for *Qest*, *QesTS*, and the null model are listed in Table I.

Table I: Mean absolute error of the null, *Qest* and *QesTS* models in predicting the test set molecular (reactant, product and transition state) or transition state partition functions on the log scale. Percent error is given compared to the test set standard deviation.

|              | Null MAE        | *Qest* MAE      | *QesTS* MAE  |
|--------------|-----------------|-----------------|--------------|
| $\log Q$     | 72.5 (34.5%)    | 4.37 (2.1%)     | N/A          |
| $\log Q_{TS}$| 72.1 (35.1%)    | 4.25 (2.1%)     | 4.01 (2.0%)  |

Given that both models predicted with a low error on the test set, we combined these to predict transition state partition functions with machine learned reactant and product partition functions. We will discuss this "Double" model in the next section. The trained models are available on GitHub.[28]

**Reaction rate constants computed with machine learned partition functions**

With our *Qest* and *QesTS* partition function estimators, and the existing activation energies,[23] we computed transition state theory reaction rate constants for 1,086 test set reactions. We recall the expression for the transition state theory,[29,30] TST, reaction rate constant, $k^{TST}(T) = \frac{1}{\beta h} \frac{Q_{TS}(T)}{Q_R(T)} e^{-\beta E_a}$, where $E_a$ is the activation energy, $\beta = 1/k_B T$ while

(*) Corresponding authors: evankomp@uw.edu, valleau@uw.edu

$Q_{TS}$ and $Q_R$ are the transition state and reactant partition functions at a given temperature $T$.

To understand how machine learned partition functions influence the accuracy of the reaction rate constant, three approaches were considered. In the first both reactant and transition state partition functions were predicted from their geometries by using *Qest*. In the second approach, the reactant and product geometries and computed ab initio partition functions were used to predict the transition state partition function with *QesTS*. This approach avoids the time demanding search for a transition state geometry. In the last approach, *Qest* was used to predict reactant and product partition functions which were used, together with their geometries, as input to *QesTS* to predict a transition state partition function. This method, hereafter referred to as the *Double* predictor, only requires knowledge of the reactant and product geometries and does not need computed partition functions or transition state geometries.

The MAE of the three methods for predicting test set transition state partition functions and TST reaction rate constants is listed in Table II. For the prediction of transition state partition functions (Table II, column 1) all models have a similar MAE. The *Double* approach is the least accurate with an MAE of 2.7% while the *QesTS* approach has the lowest MAE of 2%.

Table II: Performance of transition state partition function prediction, and TST reaction rate constant calculation using predicted partition functions for our three ML based methods (*Qest*, *QesTS* and *Double*). Here $E_a$ is the reaction activation energy. Percent error is given compared to the test set standard deviation.

|  | $\log Q_{TS}(T)$ MAE | $\log k^{TST}(T)$ MAE | Required inputs for $k(T)$ |
|---|---|---|---|
| *Qest* | 4.25 (2.1%) | 4.50 (1.7%) | Reactant, transition state structures, temperature, $E_a$ |
| *QesTS* | 4.01 (2.0%) | 4.01 (1.5%) | Reactant, product structures and partition functions, temperature, $E_a$ |
| *Double* | 5.58 (2.7%) | 4.50 (1.7%) | Reactant, product structures, temperature, $E_a$ |

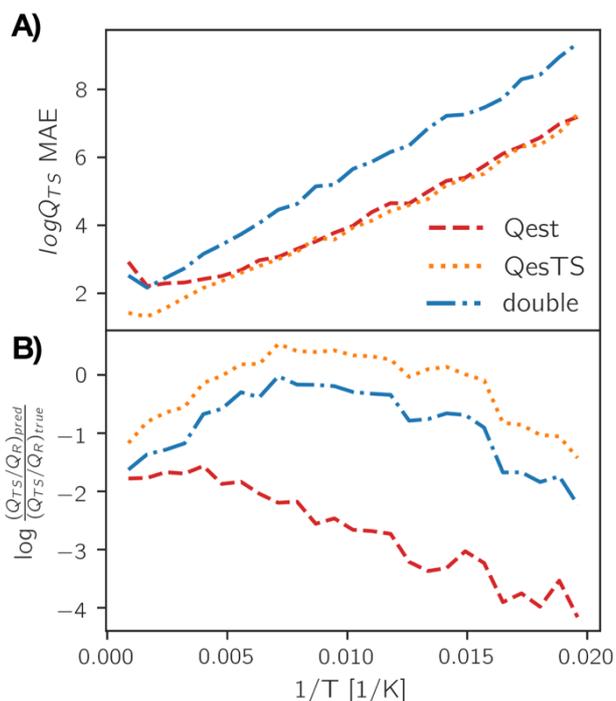

Figure 4: Panel A) MAE averaged over temperature bins on the test set for predictions of the logarithm of the transition state partition function using *Qest, QesTS,* and *Double* models. While the error of the transition state logged partition function increases as temperature decreases, the ratio of transition state to reactant partition function, which is the prefactor for TST rate constants, has low error. This is shown in the plot of the logged ratio of true

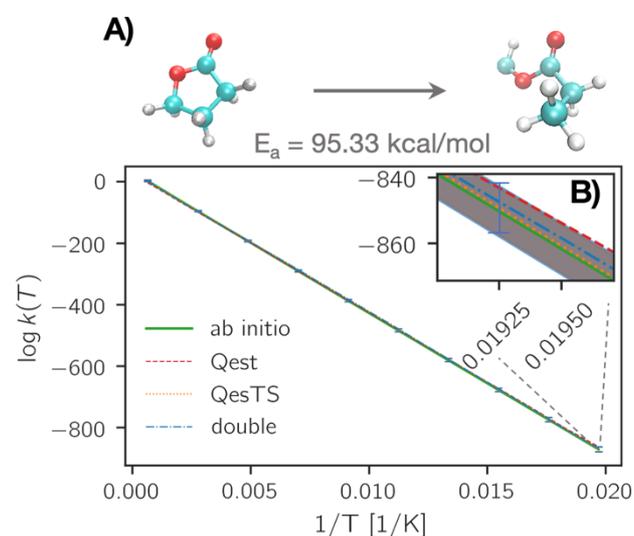

Figure 5: Panel A) Natural logarithm of the reaction rate constant for a randomly selected test set reaction: the ring breaking of gamma-Butyrolactone. Here we show the logarithm of the reaction rate constant computed with predicted *Qest* (red), *QesTS* (orange), and *Double* (blue) partition functions as well as the reaction rate constant computed using the ab initio values of the partition functions (green). The error bars correspond to the error on the entire test set averaged over temperature bins. Panel B) A zoom in on the low temperature section of the data in plot A. Here the average error is plotted as dotted margins instead of error bars. One error bar (blue) is depicted. The error on $k(T)$ for all prediction methods increases at low temperatures, however the average error is significantly smaller than the magnitude of the reaction rate constant.

The fact that *QesTS* is the best predictor is not surprising given that it was trained specifically to predict transition state partition functions while *Qest* was not. Further, *QesTS* has knowledge of reactant and product partition functions from its input. In column 2 of Table II we see that when computing reaction rate constants by using the predicted partition functions, the *Double* and *Qest* approaches have the same MAE while the *QesTS* approach remains the most accurate with an MAE of 4.01. The increase in accuracy of the *Double* approach compared to the other methods is most likely due to a cancelation of errors from *Qest* predicted reactant and *QesTS* predicted transition state partition functions. Nonetheless, we can achieve the same average accuracy of reaction rate constant calculations when using only reactant and product structures and no information on the transition state geometry. The only cost remains providing the value of the activation energy. Here it is worth noting that machine learning has successfully been employed to predict activation energies.[31,32]

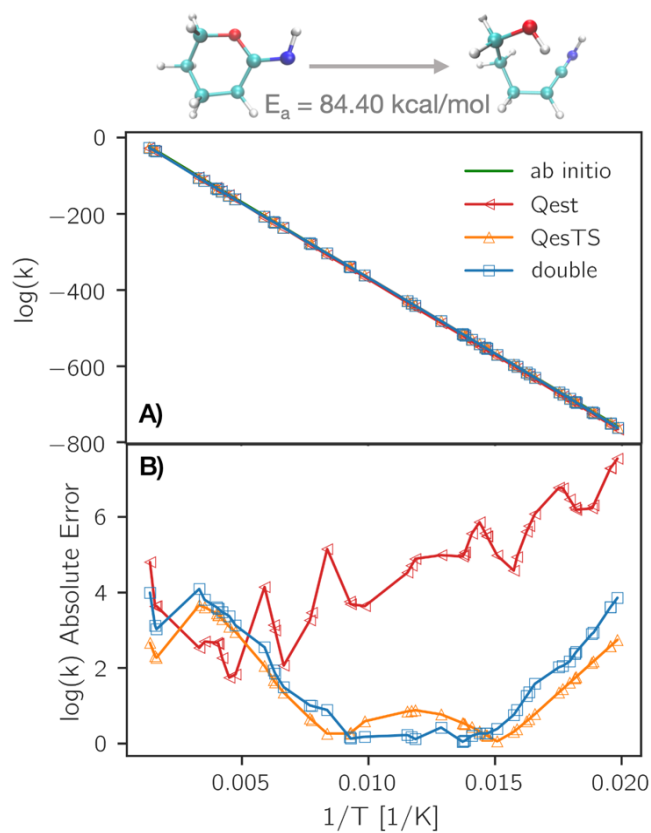

Figure 6: Panel A) Natural logarithm of the TST reaction rate constant as a function of inverse temperature, 1/T, using *Qest, QesTS,* and *Double* predicted and computed partition functions for the ring opening of tetrahydro-2H-pyran-2-imine. This reaction was randomly selected from the test set. The predicted values are all close to the ab initio values. The absolute error of these models on the logarithm of the rate constant is shown in Panel B. *Qest* prediction has larger error at low temperatures for this reaction, however all methods predict with error an order of magnitude lower than the logarithm of the rate constant.

In Figure 4, panel A we plot the error of the predicted transition state partition function with respect to the test set partition functions as a function of temperature for *Qest*, *QesTS*, and *Double*. In panel B we show the logarithm of the ratio of predicted transition state theory prefactors, $Q_{TS}/Q_R$, with respect to computed prefactors. We see some error cancelation: the error in the ratio of partition functions (panel B) is smaller than that for the single values (panel A). Errors for all models tend to increase at lower temperatures even though low temperatures were sampled frequently when generating the data. This could come from the fact that the partition function is more sensitive to small changes in temperature at low temperatures, making it more difficult to learn. Regardless, these errors do not significantly impact the predicted value of the reaction rate constant, with average error orders of magnitude lower than the value of the logarithm of the reaction rate constant. This is also seen in Figure 5, where we show the average error compared to the reaction rate constant for the ring breaking of gamma-Butyrolactone.

In Figure 6 Panel A, we show a plot of the transition state theory reaction rate constant for another reaction randomly selected from the test set, the ring opening of tetrahydro-2H-pyran-2-imine. Here the partition functions were computed using the original ab initio data (green line) and predicted with the ML models. In Panel B, we show the absolute error as the normed difference between true and predicted values. We see that the error of the models is more than an order of magnitude smaller that the value of the logarithm of the reaction rate constant which confirms the accuracy of our trained models.

## Conclusion

In this work we have computed rigid rotor, rigid body, harmonic oscillator partition functions over a broad range of temperatures for a dataset of 11,961 organic chemistry unimolecular reactions with reactant, transition state, and product structures. With this dataset two DNN based models were trained to predict partition functions at given input temperatures. *Qest* predicts a molecular partition function from the molecule's featurized geometry, *QesTS* predicts transition state partition functions from reactant and product featurized geometries and their partition functions; lastly *Double* uses *QesTS* to predict $Q_{TS}$ with *Qest* estimated reactant and product partition functions as well as reactant and product geometries. The *Double* approach requires no knowledge of the transition state structure; it only requires reactant and product structures. We showed that that these estimators are accurate in their prediction and the MAE on the logarithm of the partition function is on the order of 2% (Table I). With these predicted partition functions, we computed transition state theory reaction rate constants and found an MAE of $1.6\pm0.1\%$ on $\log k(T)$ (Table II). Predictions of individual unseen test set reactions such as the ring breaking of tetrahydro-2H-pyran-2-imine closely followed the exact test set values.

(*) Corresponding authors: evankomp@uw.edu, valleau@uw.edu

The models we have created in tandem with activation energy predictors provide an approach to predict reaction rate constants without the need to search for minimum energy paths. Our models enable a more rapid estimation of reaction dynamics in the context of coupled reactions, reaction networks, and reactor design.

In the future, we aim to move beyond unimolecular reactants and consider bimolecular reactions as well as reactions which occur in the presence of a solvent. We also plan to investigate other machine learning approaches, for instance, on employing end-to-end message passing models, given their recent success for the prediction of adjacent molecular and reaction quantities.[33,34]

## Data and Software Availability

The partition function dataset can be found on Zenodo.[25] The trained ML estimators are available on GitHub.[28]

## Conflicts of interest

There are no conflicts to declare.

## Acknowledgements

The authors would like to acknowledge the Hyak supercomputer system at the University of Washington for support of this research.

# Supplementary Information for
# "Low-cost prediction of molecular and transition state partition functions via machine learning"

Evan Komp[*] and Stéphanie Valleau[*,†]

(*) *Department of Chemical Engineering, University of Washington, Seattle, Washington 98195, United States*

† Corresponding authors: evankomp@uw.edu, valleau@uw.edu

## Table of Contents

## 1  Dataset

In **Figure S1** we show the activation energies of the reactions in the dataset used to generate the partition function dataset. In **Table S1** we show the partition function equations employed to compute $Q(T)$.

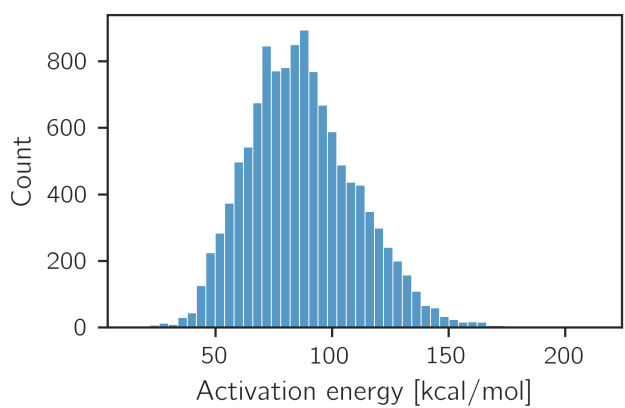

**Figure S1**: Histogram of ground state activation energies of all reaction used to generate the partition functions dataset. The average value of the activation energies $\langle E_a \rangle = 85.3$ kcal/mol is much higher than that of most organic chemistry reactions because these reactions involve breaking covalent bonds in unimolecular reactants.

**Table S1**: Equations employed to compute the translational, vibrational, and rotational partition functions with the rigid body, harmonic oscillator, and rigid rotor approximations. Here $m$ is the molecular mass, $V$ the volume, $\omega$ are the wavenumbers, $\theta_i = h^2/8\pi^2 I_i$ where $I_i$ is the moment of inertia and $\sigma$ is the symmetry number. The electronic partition function was approximated as equal to the electronic ground state degeneracy, $g_{el,0}$.

|   | Partition function |
|---|---|
| **Translational** | $Q_{trans} = V \left( \dfrac{\sqrt{2\pi m k_B T}}{h} \right)^3$ |
| **Vibrational** | $Q_{vib} = \displaystyle\prod_{i=1}^{3N-6} \dfrac{e^{-\hbar\omega_i/2k_B T}}{1 - e^{-\frac{\hbar\omega_i}{k_B T}}}$ |
| **Rotational** | $Q_{rot} = \dfrac{\sqrt{\pi}}{\sigma} \sqrt{\dfrac{(k_B T)^3}{\theta_x \theta_y \theta_z}}$    3D structure <br> $Q_{rot} = \dfrac{k_B T}{\sigma \theta}$    linear structure |

## 2 Featurization

### 2.1 Search for optimal input features

For the *Qest* model, we screened over a series of geometry based input features which included Autocorrelation (AC) [1], EncodedBonds (EB) [2], and Coulomb Matrix (CM)[3] as well as Smooth Overlap of Atomic Positions (SOAP) [4], and Many Body Tensor Representation (MBTR) [5]. To create these, the MolML [2] and DScribe [6] software packages were used. See section 2.2. Graph based featurization techniques were not considered and this is in plan for future work. For each featurization we tested feature and target standardization; we considered min-max scaling and gaussian normalization. The model hyperparameters used during this screening are shown in **Table S2**.

Overall, we searched over 45 possible combinations of input feature, feature standardization, and target standardization and found that EncodedBonds with feature min-max scaling and target normalization were optimal. The average performance of the different featurizers is shown in **Figure S2**. We repeated the screening of standardization for *QesTS* using EncodedBonds, which yielded no standardization as optimal.

**Table S2**: Hyperparameters used during data screening.

| Hyperparameter | Value |
|---|---|
| layer sizes | [200, 200] |
| learning rate | 0.001 |
| batch size | 100 |
| epochs | 20 |
| hidden activation function | relu |
| l2 regularization | 0.0 |
| bias initialization constant | 1.0 |
| weight initialization stdev | 0.2 |
| loss | mean squared error |

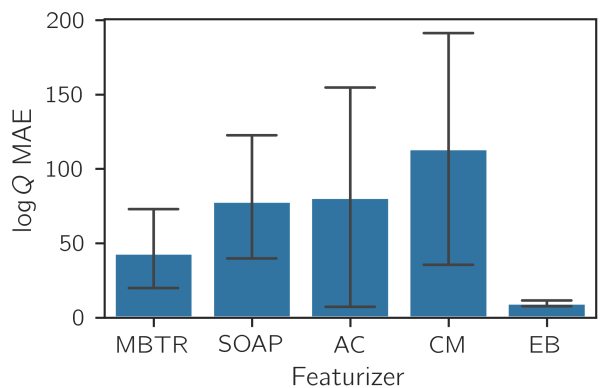

**Figure S2**: Average performance of each featurizer tested during screening. Bars represent the cross validated mean absolute error, while error bars are the standard deviation across the data standardization combinations tested for each featurizer. Models using Encoded Bonds performed best on average. Also, the best performing combination used Encoded Bonds with feature min-max scaling and target normalization.

## 2.2 Featurization parameters

The parameters used to produce the input features are described in subsection 2.2.1 and 2.2.2. Smooth Overlap of Atomic Positions (SOAP) and Many-body Tensor Representation (MBTR) input features were produced with the DScribe software, while Autocorrelation, Encoded Bonds, and Coulomb Matrix were produced with the MolML software.

### 2.2.1 DScribe Software

The DScribe software package[6] was used to generate SOAP and MBTR input features. The parameters used are listed below.

*SOAP – Smooth Overlap of Atomic Positions*
The parameters used to produce SOAP features are listed in python dictionary format:
{
   'species': ["H", "C", "O", "N"],
   'rcut': 6.0,
   'nmax': 8,
   'lmax': 6,

```
    'average': 'inner'
}
```
For a description of these parameters, we refer the reader to Ref [6].

*MBTR – Many-body Tensor Representation*

The parameters used to produce MBTR features are listed hereafter in python dictionary format:
```
{
    species=["H", "O", 'C', 'N'],
    k1={
        "geometry": {"function": "atomic\_number"},
        "grid": {"min": 0, "max": 8, "n": 100, "sigma": 0.1},
    },
    k2={
    "geometry": {"function": "inverse\_distance"},
    "grid": {"min": 0, "max": 1, "n": 100, "sigma": 0.1},
    "weighting": {"function": "exponential", "scale": 0.5, "cutoff": 1e-3},
    },
    k3={
    "geometry": {"function": "cosine"},
    "grid": {"min": -1, "max": 1, "n": 100, "sigma": 0.1},
    "weighting": {"function": "exponential", "scale": 0.5, "cutoff": 1e-3},
    },
    periodic=False,
    normalization='none',
    flatten=True
}
```

We refer the reader to Ref [6] for a description of these parameters.

### 2.2.2 MolML Software

MolML [2] was used to define the EncodedBonds, Coulomb Matrix and Autocorrelation features. MolML determines parameters such as the element types present and the maximum number of atoms during featurization. Due to limited RAM, the entire dataset could not be loaded at once, so it was scanned for a set of 3 structures which contained both the full array of chemical elements in the dataset, as well as the largest molecular system. The featurizers were fit to this minimal set, and then used to transform the entire dataset in chunks. The minimum subset of systems is given below in xyz format and angstrom units:

| | | | |
|---|---|---|---|
| C | -2.301345240000000 | 0.245555560000000 | -0.146458820000000 |
| O | -1.801117640000000 | -1.068638690000000 | -0.132535600000000 |
| C | -0.698207700000000 | -1.254313550000000 | -0.983955260000000 |
| C | 0.498785910000000 | -0.503501480000000 | -0.581849450000000 |
| C | 1.474872740000000 | 0.110834400000000 | -0.254355130000000 |
| C | 2.665120440000000 | 0.852752900000000 | 0.143128950000000 |
| H | -2.602697440000000 | 0.544988140000000 | -1.159271040000000 |
| H | -3.175412910000000 | 0.262006730000000 | 0.502576870000000 |
| H | -1.561197990000000 | 0.962926400000000 | 0.225284290000000 |
| H | -0.479999560000000 | -2.323300650000000 | -0.969310430000000 |
| H | -0.961290100000000 | -0.982075860000000 | -2.016315420000000 |
| H | 3.153957870000000 | 0.378701800000000 | 0.995343110000000 |
| H | 3.380779340000000 | 0.898604960000000 | -0.679097780000000 |
| H | 2.407752320000000 | 1.875459330000000 | 0.422868020000000 |
| O | 2.288293630000000 | 0.484583070000000 | -0.121712670000000 |
| C | 1.127501880000000 | 0.193607830000000 | -0.071145460000000 |
| C | -0.184064230000000 | 0.924152480000000 | 0.259412420000000 |
| C | -0.873748740000000 | -0.432000910000000 | -0.015729600000000 |
| N | 0.451962880000000 | -0.971798030000000 | -0.300035950000000 |
| H | -0.242967130000000 | 1.286299400000000 | 1.283947400000000 |
| H | -0.427416880000000 | 1.726023410000000 | -0.434781960000000 |
| H | -1.362208720000000 | -0.885687600000000 | 0.848217390000000 |
| H | -1.550559340000000 | -0.441221400000000 | -0.871782480000000 |
| H | 0.773206630000000 | -1.883958250000000 | -0.576389100000000 |

23

| | | | |
|---|---|---|---|
| C | 2.349176060000000 | 0.084494630000000 | -0.905389770000000 |
| C | 1.627675400000000 | 0.053488250000000 | 0.436342020000000 |
| C | 0.236369270000000 | -0.590909200000000 | 0.414770920000000 |
| C | -0.342881300000000 | -0.607355670000000 | 1.828561880000000 |
| C | -0.712898880000000 | 0.033731390000000 | -0.629030880000000 |
| C | -0.976517400000000 | 1.521383910000000 | -0.403857370000000 |
| C | -2.030065150000000 | -0.732800180000000 | -0.733683520000000 |
| H | 2.385265370000000 | -0.910486730000000 | -1.357281740000000 |
| H | 3.377237770000000 | 0.431331870000000 | -0.786465570000000 |
| H | 1.861357880000000 | 0.752318050000000 | -1.618183410000000 |
| H | 2.239225610000000 | -0.501941370000000 | 1.154397840000000 |
| H | 1.554048780000000 | 1.069878880000000 | 0.838340710000000 |
| H | 0.375865430000000 | -1.636038860000000 | 0.106313200000000 |
| H | 0.341081900000000 | -1.112555420000000 | 2.513879630000000 |
| H | -0.495666310000000 | 0.405190570000000 | 2.210740010000000 |
| H | -1.298151790000000 | -1.131761590000000 | 1.876033720000000 |
| H | -0.213934640000000 | -0.065520660000000 | -1.599015070000000 |
| H | -1.615287910000000 | 1.919106830000000 | -1.195418250000000 |
| H | -1.487633380000000 | 1.698516490000000 | 0.545714450000000 |
| H | -0.054749190000000 | 2.106284580000000 | -0.401858700000000 |
| H | -2.609548020000000 | -0.389412500000000 | -1.593433690000000 |
| H | -1.858128490000000 | -1.805335340000000 | -0.854936010000000 |
| H | -2.651841000000000 | -0.591607940000000 | 0.153459620000000 |

# 3  Optimization of hyperparameters

After screening for input features and standardization (Section S2) we carried a hyperparameter optimization. The search space for hyperparameter optimization is shown in **Table S3**.
The Tree Parzen [7,8] sampler along with the median pruning algorithm was used to identify the optimal hyperparameters (see **Table S4**) with Optuna [9] using a 5 trial startup for both models. This corresponds to 1258 and 7685 completed trials for the *Qest* and *QesTS* model, respectively.
The number of epochs over which to train the model for the optimal hyperparameters was determined by using early stopping averaged over the 5 folds. From this we found that 39 and 27 epochs were optimal stopping points for the *Qest* and *QesTS* models respectively.

**Table S3**: Parameter space tested for hyperparameter optimization. Layer parameters such as bias initialization are applied to all hidden layers uniformly; the search space is restricted by not considering layers having different parameters. One to three total hidden layers are tested, and in each case all layers have the same number of neurons chosen from the range [2 – 2000]. MSE loss is used.

| Hyperparameter | Search space | Distribution |
|---|---|---|
| layer sizes | 1-3 layers, 2 - 2000 neurons | categorical choice |
| learning rate | (1E-5, 1E-1) | loguniform |
| batch size | (32, 8000) | uniform integer |
| hidden activation function | relu, tanh, softsign, sigmoid, softmax | categorical choice |
| l2 regularization | (0.0, 0.1) | uniform |
| bias initialization constant | (-1.0, 1.0) | uniform |
| weight initialization stdev | (1E-3, 1E+0) | loguniform |

**Table S4**: Optimum hyperparameters for both the Qest and QesTS DNN models found from TPE hyperparameter optimizations using Encoded Bonds as input features.

| Hyperparameter | Qest DNN | QesTS DNN |
|---|---|---|
| layer sizes | [816, 816] | [749, 749] |
| learning rate | 1.58E-4 | 3.13E-4 |
| batch size | 34 | 2609 |
| hidden activation | relu | relu |
| l2 regularization | 8.68E-6 | 4.20E-5 |
| bias initial value | -0.664 | -0.207 |
| weight initialization stdev | 0.407 | 0.059 |

# 4 Outliers

To limit extrapolation in the final model, outliers were dropped from the overall dataset as following. We identified reactions containing any example with 12 or more feature values $\pm 6$ standard deviations away from that feature's mean. This resulted in removing 1,108 reactions.